\documentclass{ws-procs10x7}

\begin{document}

\title{
  Recent results from lattice calculations
}
\author{Shoji Hashimoto}
\address{
  High Energy Accelerator Research Organization (KEK),
  Tsukuba 305-0801, Japan\\
  e-mail: shoji.hashimoto@kek.jp
}

\twocolumn[\maketitle\abstract{
  Recent results from lattice QCD calculations relevant
  to particle physics phenomenology are reviewed.
  They include the calculations of strong coupling constant,
  quark masses, kaon matrix elements, and $D$ and $B$ meson
  matrix elements.
  Special emphasis is on the recent progress in the
  simulations including dynamical quarks.
}]

\section{Introduction}
Since it was invented by K. Wilson\cite{Wilson:1974sk} in
1974, lattice QCD has grown into an important tool for
the analysis of low energy
regime of QCD, where the non-perturbative dynamics of quarks
and gluons becomes essential. 
In this talk I review the status of the calculation of
several quantities relevant to particle physics phenomenology, 
such as the determination of fundamental parameters of QCD,
and the hadron matrix elements of $K$, $D$ and $B$ mesons.

Although lattice gauge theory starts with first principles
in QCD calculations, due to the limitation of computational
resources one has to resort to
several approximations, such as finite lattice spacing $a$, 
finite spatial extent $L$ of the lattice, and relatively
large quark mass $m_q$.
They introduce sources of systematic uncertainties, which
can in principle be reduced by performing extrapolations to
the appropriate limits.

Among other approximations, the most serious one was the
quenched approximation, in which gluon vacuum polarization
of by quarks is neglected.
It reduces the computational cost by orders of magnitude and
so has been used for most lattice calculations, but
the systematic uncertainty is not under any quantitative
control. 
The most important development since the last ICHEP
conference in 2002, where Lellouch summarized the lattice
results\cite{Lellouch:2002nj}, is that the simulations
beyond the quenched approximation---unquenched QCD--- have
been carried out for many important quantities. 

In this review, I firstdiscuss the theoretical issues in 
dynamical quark simulations, which are related to
formulations and algorithms for fermions on the lattice.
Then I cover several applications, including the
determination of the strong coupling and quark masses,
and calculations of weak matrix elements.
Other reviews of recent results can be found in
\cite{Hashimoto:2004fv,Wingate:2004xa}.

\section{Issues in dynamical QCD}
Since dynamical fermion simulations involve many inversions
of fermion matrix, it is substantially harder 
to simulate light quarks at their physical mass values.
Therefore, an extrapolation, called the chiral
extrapolation, in the light quark mass from feasible quark
masses to the physical up and down quark masses is necessary.
The extrapolation is best controlled with the functional
form predicted by the chiral perturbation theory (ChPT),
which provides a systematic expansion of low energy QCD for
small quark masses. 

The question is then whether the region of quark masses in
lattice simulations has enough overlap with the convergence
radius of ChPT.
For the pion decay constant, for instance, at NLO ChPT
predicts a non-analytic behavior $-m_\pi^2\ln m_\pi^2$,
called the chiral log.
The result from the JLQCD collaboration
\cite{Hashimoto:2002vi,Aoki:2002uc} clearly shows that there
is no signal of curvature for the ``pion'' mass above
550~MeV, which is currently the lowest available pion mass
with (improved) Wilson fermion formulations.
More recently the MILC collaboration\cite{Aubin:2004fs} has
performed a simulation with the pion mass as low as 250~MeV
using the (improved) staggered fermions.
They found that the chiral log shows up below 500~MeV,
showing a clear advantage of the staggered fermion.
In fact, using the simulations including 2+1 (up, down and
strange) flavors of staggered quarks, it has been
demonstrated that several fundamental physical quantities
are in good agreement with the experiment\cite{Davies:2003ik}. 

Wilson fermions explicitly violate chiral symmetry at finite
lattice spacing, and the massless quark 
limit has to be reached by tuning a mass parameter.
Because the massless point itself fluctuates statistically
in the Monte Carlo simulation, the singularity of the
massless limit could show up earlier in the chiral
extrapolation, and the dynamical fermion simulation becomes
much harder near the chiral limit than expected from naive
scaling law.
With the staggered fermion, on the other hand, there is an
exact U(1) chiral symmetry and the massless limit is fixed.

The price one has to pay for the staggered fermion is the
complication of species doubling.
The formulation necessarily involves four species
(or {\it tastes}) of fermions.
They mix among themselves at finite lattice spacing.
As a result, there are 16 pions on the lattice, only one of
which becomes massless in the chiral limit.
All other hadrons are also duplicated, and such effects
could become a source of systematic uncertainty.

More seriously, one has to take a square-root or fourth-root
of the fermion determinant in order to represent degenerate
up and down quarks or slightly heavier strange quark using
the four-taste staggered fermion.
Any local Dirac operator corresponding to the
fourth-rooted staggered fermion determinant has not been
found so far\footnote{ 
  It was shown that the square-root of the staggered
  operator is {\it non}-local\cite{Bunk:2004br,Hart:2004sz}.
}.
Without the locality one cannot prove the universality, 
{\it i.e.} the continuum limit is {\it the} QCD.
Therefore, it is potentially a fundamental problem of the
calculations employing the staggered fermions\footnote{
  There is also a positive indication, {\it i.e.}
  the eigenvalue spectrum of the staggered operator shows an
  approximate four-fold
  degeneracy\cite{Follana:2004sz,Durr:2004as}.
}.

Theoretically, the best solution is to employ the fermions
satisfying the Ginsparg-Wilson relation, which have an
exact chiral symmetry at finite lattice spacing without
sacrificing the flavor symmetry\cite{Luscher:1998pq}.
Neuberger (overlap)
fermions\cite{Neuberger:1997fp,Neuberger:1998wv} and  
domain-wall
fermions\cite{Kaplan:1992bt,Shamir:1993zy,Furman:1994ky} fall
into this class. 
Exploratory dynamical simulations using the domain-wall
fermion have already been performed\cite{Dawson:2003ph}.

In the following discussion I assume that the fourth-rooted
staggered fermion is a correct, or at least effective,
description of QCD and do not quote errors associated with
it.

\section{Fundamental QCD parameters}

\subsection{Strong coupling constant}
The strong coupling constant $\alpha_s(M_Z)$ can be
determined using lattice QCD by converting
$\alpha_s^{lat}(1/a)$ to the continuum definition 
$\alpha_s^{\overline{MS}}(M_Z)$ by perturbation theory. 
The quarkonium spectrum is often used to set $1/a$,
because it is insensitive to other systematic errors, such
as those from finite volume and chiral extrapolation of
light quarks. 
To improve the perturbative expansion the renormalized
coupling\cite{Lepage:1992xa} $\alpha_V(q^*)$, which is
defined through the heavy quark potential, is used with an
appropriate scale $q^*$.

\begin{figure}[tb]
  \centering
  \includegraphics*[width=6.2cm]{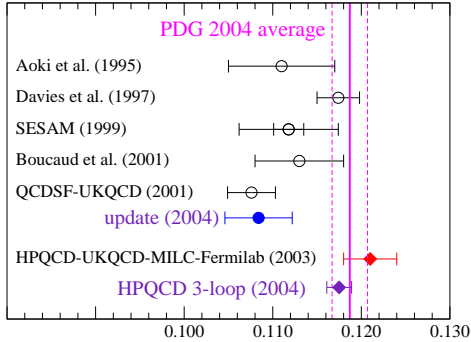}
  \caption{
    Strong coupling constant $\alpha_s(M_Z)$.
     Lattice data are from 
     Aoki et al.\protect\cite{Aoki:1994pc},
     Davies et al.\protect\cite{Davies:1997mg},
     SESAM\protect\cite{Spitz:1999tu},
     QCDSF-UKQCD\protect\cite{Booth:2001qp} and its
     update\protect\cite{Gockeler:2004ad},
     HPQCD-UKQCD-MILC-Fermilab\protect\cite{Davies:2003ik} and its
     update\protect\cite{Mason_lat04}.
  }
  \label{fig:alpha_s}
\end{figure}

Figure~\ref{fig:alpha_s} summarizes lattice results
using unquenched simulations.
After including the dynamical quark effects, the most
important source of systematic error is the unknown higher
order perturbation theory.
Previous results used two-loop matching, but
this year the HPQCD collaboration\cite{Mason_lat04} has
carried out three-loop calculations and reported a
(preliminary) result with a substantially reduced error, 
$\alpha_s(M_Z)$ = 0.1175(15), which is in good agreement
with the PDG 2004 average 0.1182(20)\cite{Eidelman:2004wy}. 
They use the simulation data with 2+1 flavors of improved
staggered fermions at three lattice spacings and confirm
that various input quantities to determine $\alpha_V(q^*)$
give a consistent result within estimated four-loop errors.
It should be noted that the results with the Wilson-type
fermion\cite{Booth:2001qp,Gockeler:2004ad} is significantly
lower.
It is likely an unknown higher order effect, but is not
fully understood. 

\subsection{Light quark masses}
The light (up, down and strange) quark masses are determined
from the pion and kaon masses, 
{\it e.g.}, using the PCAC relation such as
$m_K^2=B(\bar{m}+m_s)$ at the leading order of $m_q$.
$\bar{m}$ denotes an average up and down quark mass and
$m_s$ is the strange quark mass.
Since the effect of chiral log is not significant for
$m_s$, most of the calculations use a linear fit in an
average light quark mass for interpolation.
For the calculation of $\bar{m}$, on the other hand, the NLO
effect of ChPT is important. 

Because $m_q$ is regularization dependent, the lattice
results are usually quoted in the $\overline{\mathrm{MS}}$
scheme by using perturbative matching and sometimes also using 
non-perturbative techniques at intermediate steps. 
For the case of the Wilson-type fermions, the determination
through axial Ward identity (AWI) and vector Ward identity
(VWI) could be different at finite lattice spacing\footnote{
  For further discussions on the quark mass calculations I
  refer to \cite{Wittig:2002ux,Gupta:2003fn}. 
}.

\begin{figure}[tb]
  \centering
  \includegraphics*[width=6.2cm]{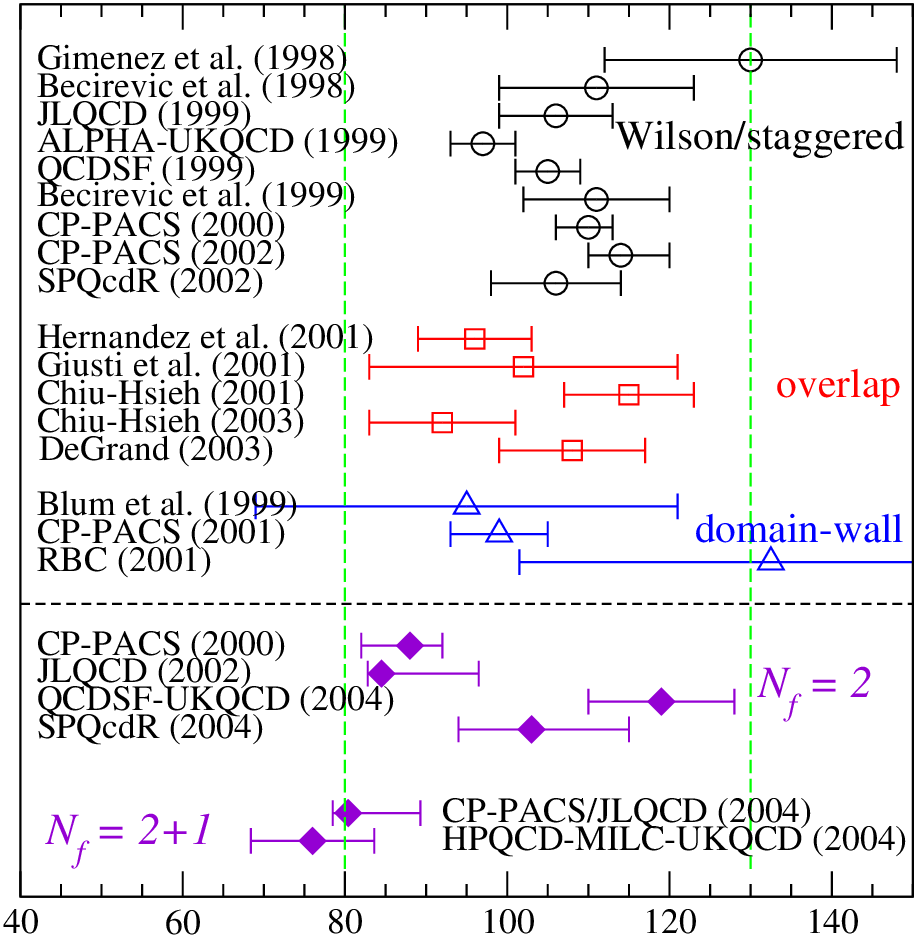}
  \caption{
    Strange quark mass $m_s(2\,\mathrm{GeV})$ (MeV) from
    lattice QCD.
    Both quenched and unquenched results are listed.
    Quenched results (upper panel) are from
    Gimenez et al.\protect\cite{Gimenez:1998uv},
    Becirevic et al.\protect\cite{Becirevic:1998yg},
    JLQCD\protect\cite{Aoki:1999mr},
    ALPHA-UKQCD\protect\cite{Garden:1999fg},
    QCDSF\protect\cite{Gockeler:1999cy},
    Becirevic et al.\protect\cite{Becirevic:1999kb},
    CP-PACS\protect\cite{AliKhan:2000mw,Aoki:2002fd},
    SPQcdR\protect\cite{Becirevic:2002jg},
    Hernandez et al.\protect\cite{Hernandez:2001hq},
    Giusti et al.\protect\cite{Giusti:2001pk},
    Chiu-Hsieh\protect\cite{Chiu:2002rk,Chiu:2003iw},
    DeGrand\protect\cite{DeGrand:2003in},
    Blum et al.\protect\cite{Blum:1999xi},
    CP-PACS\protect\cite{AliKhan:2001wr},
    RBC\protect\cite{Dawson:2002nr}.
    Unquenched results are from 
    CP-PACS\protect\cite{AliKhan:2000mw},
    JLQCD\protect\cite{Aoki:2002uc},
    QCDSF-UKQCD\protect\cite{Gockeler:2004rp},
    SPQcdR\protect\cite{Becirevic:2004sb},
    CP-PACS/JLQCD\protect\cite{Ishikawa:2004xq},
    HPQCD-MILC-UKQCD\protect\cite{Aubin:2004ck,Aubin:2004fs}.
    PDG 2004\protect\cite{Eidelman:2004wy} average is shown by a
    dashed band.
  }
  \label{fig:ms}
\end{figure}

In Figure~\ref{fig:ms}, I compile the lattice results for
strange quark mass for both quenched and unquenched
calculations.
In the quenched approximation, systematic studies of 
the non-perturbative matching and the continuum
extrapolation have been extensively studied and the results
are in agreement within $\approx$ 10\% quenching error,
which appears as a dependence on the input quantity to set
the lattice scale, 
{\it e.g.} $m_\rho$, $f_K$, etc.

The CP-PACS\cite{AliKhan:2000mw} and JLQCD\cite{Aoki:2002uc}
collaborations found that $m_s(2\mbox{~GeV})$ becomes
significantly lower by the effects of two dynamical
flavors.\footnote{
  In these works, the matching is done perturbatively.
  The QCDSF-UKQCD collaboration\cite{Gockeler:2004rp}
  calculated the non-perturbative matching factor for the
  VWI determination, and found it larger by about 20\% than
  the one-loop estimate.
  The central values of \cite{AliKhan:2000mw,Aoki:2002uc}
  are taken from the AWI definition, however. 
}
Recently, 2+1-flavor calculations are reported by
the CP-PACS/JLQCD\cite{Ishikawa:2004xq} and
HPQCD-MILC-UKQCD\cite{Aubin:2004ck,Aubin:2004fs}
collaborations. 
Their results are consistent with each other and slightly
lower than the two-flavor data.  
My average is $m_s(2\mbox{~GeV})=78\pm 10$~MeV.

Determination of light quark mass $\bar{m}$ or the ratio
$m_s/\bar{m}$ is sensitive to the chiral extrapolation.
At the leading order of $m_q$ the ratio is given by the
physical meson masses as $m_K^2/m_\pi^2-1$ = 25.9, and a
NLO ChPT analysis yields
24.4$\pm$1.5\cite{Leutwyler:1996qg}.
The lattice calculation can be used to improve this
estimate. 
The small quark mass reached by the MILC simulation enabled 
them to include NLO ChPT terms in the
fit\cite{Aubin:2004ck,Aubin:2004fs} as well 
as the correction terms to describe the taste symmetry
breaking\cite{Aubin:2003mg} and higher order effects.
Their result 27.4$\pm$4.2 is consistent with the NLO ChPT
analysis but slightly higher, suggesting non-negligible
higher order effect.

\subsection{Heavy quark masses}
The charm quark is not too heavy to describe with the 
$O(a)$-improved Wilson fermion action adopting the
naive estimate of discretization effect $O((am_c)^2)$.
It can in principle be eliminated by taking the continuum
limit, which is feasible in the quenched approximation
and precise results $\bar{m}_c(\bar{m}_c)$ = 
1.30(3)\cite{Rolf:2002gu} and
1.32(3)\cite{deDivitiis:2003iy}~GeV are obtained.
If one takes the non-relativistic dynamics of charm quark
inside the $D_{(s)}$ meson into
account\cite{El-Khadra:1996mp,Kronfeld:2000ck},
the discretization error is not as large as
$O((am_c)^2)$.
Recent work indicates that the discretization effect is much
smaller\cite{Dougall:2004hx}, and an unquenched calculation
is being performed.

For the bottom quark the conventional approach 
fails for the lattice scale $1/a\sim$ 2--3~GeV.
Instead, the heavy quark effective theory (HQET) is a
good approximation up to corrections of order
$O(\Lambda_{\mathrm{QCD}}^2/m_b)\simeq$ 30~MeV.
Higher order perturbation theory is essential for the
matching of $m_b$ in order to avoid large corrections due to
power divergences.
The two-loop calculation was done sometime
ago\cite{Martinelli:1998vt} and the three-loop calculation 
has been performed recently\cite{DiRenzo:2004xn}, reducing
the error to the 40~MeV level.  
Available two-flavor QCD calculations combined with the
two-loop matching yield
$\bar{m}_b(\bar{m}_b)=$
4.21(7)~MeV\cite{Gimenez:2000cj,DiRenzo:2004xn} and
4.25(11)~MeV\cite{McNeile:2004cb}.
For the latter, carefully estimated uncertainties in the
lattice scale and strange quark mass dominate the error
bar, which is expected to be reduced by 2+1-flavor
calculations.

Recently, a non-perturbative method to match HQET onto QCD
has been formulated and tested on quenched
lattices\cite{Heitger:2003nj}.
Another method to calculate $b$ quark mass without recourse
to HQET has also been proposed\cite{deDivitiis:2003iy}.
These methods may enable us to further reduce the systematic
error.

\section{Kaon physics}

\subsection{Determination of $|V_{us}|$}
The best known method to determine $|V_{us}|$, or the
Cabibbo angle, is to use the semi-leptonic $K_{l3}$ decays. 
The relevant form factor $f_+(0)$ is normalized to
one in the SU(3) limit ($\bar{m}=m_s$), and the correction
starts at the second order in 
$m_s-\bar{m}$\cite{Ademollo:1964sr}.
Calculation of the correction in a quark model yielded 
$f_+(0)$ = 0.961(8)\cite{Leutwyler:1984je}. 
Further improvement requires non-perturbative method to
calculate $f_+(0)$, and first quenched lattice calculation
has been done recently\cite{Becirevic:2004ya} using double
ratios as in the $|V_{cb}|$
calculation\cite{Hashimoto:1999yp,Hashimoto:2001nb}. 
They reported 0.960(5)(7).

$|V_{us}|$ can also be determined through the leptonic decay
$K^\pm\to\mu^\pm\nu_\mu$, once the decay constant $f_K$ is
known theoretically.
This has been attempted\cite{Marciano:2004uf} using the
recent MILC result $f_K/f_\pi$ =
1.210(4)(13)\cite{Aubin:2004fs}.
It is notable that the error is now comparable to the
semi-leptonic determination, {\it i.e.} $\sim$ 1\%, and the
result for $|V_{us}|$ is consistent.

\subsection{Kaon $B$ parameter}
Calculation of $B_K$ is one of the major goals of lattice QCD 
calculations.
$B_K$ is a matrix element of a $\Delta S=2$ four-quark
operator of $VV+AA$ chiral structure sandwiched by $K$ and
anti-$K$ states.
In the lattice calculation its chiral structure must be
maintained in order to avoid large contamination from wrong
chirality operators.
Therefore, most lattice calculations have been done using
fermion formulations which respect chiral symmetry.

\begin{figure}[tb]
  \centering
  \includegraphics*[width=6.0cm]{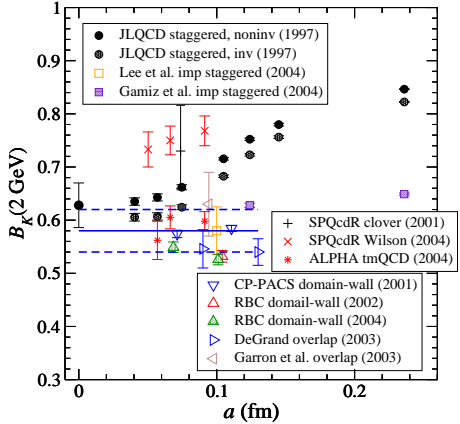}
  \caption{
    Quenched lattice results for $B_K(2\,\mathrm{GeV})$
    versus lattice spacing $a$.
    Results with staggered fermion 
    (JLQCD\protect\cite{Aoki:1997nr}, 
    Lee et al.\protect\cite{Lee:2004qv},
    Gamiz et al.\protect\cite{Gamiz:2004qx}),
    Wilson-type fermions
    (SPQcdR clover\protect\cite{Becirevic:2000ki},
    SPQcdR Wilson\protect\cite{Becirevic:2004aj})
    twisted mass fermion\protect\cite{Dimopoulos:2004xc},
    domain-wall fermion,
    (CP-PACS\protect\cite{AliKhan:2001wr},
    RBC\protect\cite{Blum:2001xb,Noaki:2004sd}), and
    overlap fermion
    (DeGrand\protect\cite{DeGrand:2003in},
    Garron et al.\protect\cite{Garron:2003cb}).
    Horizontal lines show my average.
}
  \label{fig:B_K}
\end{figure}

In the quenched approximation, the ``benchmark'' result was
given by the JLQCD collaboration\cite{Aoki:1997nr} in 1997
using (unimproved) staggered fermion:
$B_K^{N_f=0}(2\,\mathrm{GeV})$ = 0.63(4).
As shown in Figure~\ref{fig:B_K} they calculated $B_K$ at
several values of lattice spacing down to $\sim$ 0.04~fm and
extrapolated to the continuum limit assuming theoretically
expected discretization effects.
More recent improved staggered fermion
results\cite{Lee:2004qv,Gamiz:2004qx} are consistent with
the JLQCD's continuum limit already at large lattice
spacings $a\simeq$ 0.1--0.2~fm.

In the past few years, many calculations have been performed
using the Ginsparg-Wilson fermions (domain-wall and overlap)
\cite{AliKhan:2001wr,Blum:2001xb,Noaki:2004sd,%
  DeGrand:2003in,Garron:2003cb}.
The results are shown by triangles in Figure~\ref{fig:B_K}.
They are slightly lower than the JLQCD's continuum limit,
and discretization error is substantially reduced.
For some of these, the renormalization factor is computed
non-perturbatively. 

Another method to protect the lattice operator from operator
mixing is provided by chirally twisted mass lattice
QCD\cite{Frezzotti:2004wz}.
The numerical result for $B_K$\cite{Dimopoulos:2004xc} is
almost flat in $a$ and consistent with the Ginsparg-Wilson
fermion results\footnote{
  A related proposal has been made for Wilson
  fermions\cite{Becirevic:2000cy} to avoid the wrong
  chirality operators.
  A numeical result\cite{Becirevic:2004aj} suggests large
  $O(a)$ discretization effect for unimporved Wilson
  fermion. 
}. 

In view of these improved calculations, which have smaller
discretization effect and non-perturbative renormalization, 
I recommend a slightly lower value for a quenched world
average, $B_K^{N_f=0}(2\,\mathrm{GeV})$ = 0.58(4), which is
shown in Figure~\ref{fig:B_K} by horizontal lines.

Dynamical quark effects were previously estimated from 
unimproved staggered
simulations\cite{Ishizuka:1993ya,Kilcup:1993pa,Lee:1996yg,Kilcup:1996hp} 
as only slightly negative\cite{Soni:1995qq} or
positive\cite{Sharpe:1998hh}.
Due to the large discretization errors inherent in the
unimproved staggered fermion, it was difficult to
disentangle the sea quark effect from the discretization
effect. 
This year, new unquenched calculations have appeared using
the $O(a)$-improved Wilson fermion\cite{Flynn:2004au},
improved staggered fermion\cite{Gamiz:2004qx}, and
domain-wall fermion\cite{Dawson:2004gu}.

\begin{figure}[tb]
  \centering
  \includegraphics*[width=6.0cm]{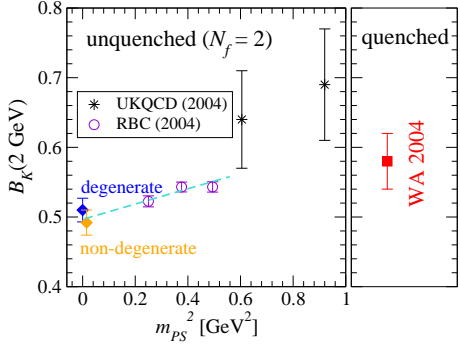}
  \caption{
    Unquenched $B_K$.
    Data from UKQCD\protect\cite{Flynn:2004au} and
    RBC\protect\cite{Dawson:2004gu} are shown as a function
    of pion mass squared.
    The right panel shows quenched world average.
  }
  \label{fig:B_K_unq}
\end{figure}

Figure~\ref{fig:B_K_unq} shows sea quark mass (or mass of
pion composed by sea quarks) dependence of $B_K$.
The RBC calculation with dynamical domain-wall
fermion\cite{Dawson:2004gu} is very precise and
shows a tread to decrease toward the chiral limit.
The SU(3) breaking $m_s\neq m_d$ effect is estimated to be
about $-$3\%.
Since the result is still preliminary and the slope
in the chiral extrapolation seems to rely on a single point
at $m_{PS}^2\simeq$ 0.25~GeV$^2$, I take this as an
indication of quenching error rather than a central value of
unquenched QCD, and recommend an average
$B_K(2\,\mathrm{GeV})$ = 0.58(4)($^{+0}_{-9}$), where the
second asymmetric error reflects the quenching error.

\section{Heavy quarks}

\subsection{$D$ meson decays}
The CLEO-c and BES III experiments promise to measure the
$D_{(s)}$ decays at a few \% accuracy, which provides a
stringent test of lattice simulation of
heavy quarks.
The most relevant hadronic quantities to be calculated on
the lattice are the leptonic decay constants and
semi-leptonic decay form factors.

As in the charm quark mass calculation the
large discretization effect of $O((am_c)^2)$ must be
eliminated for precise calculation of $D_{(s)}$ meson decay
constants, unless one uses effective theory approaches.
Continuum extrapolation has recently been
performed in the quenched approximation, yielding 
$f_{D_s}$ = 252(9)~MeV\cite{Juttner:2003ns} and
240(5)(5)~MeV\cite{deDivitiis:2003wy}.
They are consistent with the previous world average
230(14)~MeV\cite{Ryan:2001ej} from the 
calculations using the Fermilab heavy quark
formulation\cite{El-Khadra:1996mp}, which applies the
idea of HQET for the Wilson-type lattice fermion. 

The Fermilab-MILC-HPQCD collaboration has carried out
calculations of $f_D$ and $f_{D_s}$ in the presence of
2+1-flavors of staggered sea quarks\cite{Simone:2004fr}.
Their result is $f_{D_s}$ = 263($^{+5}_{-9}$)(24)~MeV.
The first error is statistical, and the second error
reflects their estimate of systematic error due to matching 
of the heavy quark action and current to QCD (7\%),
discretization effects of the light quark (4\%), charm quark
mass determination (4\%), etc. 
To improve the heavy quark matching, one requires
perturbative calculation for complicated lattice actions
including many $O(1/m_Q)$ terms, and such work is in
progress using automated perturbative calculation
technique\cite{Nobes:2003nc}. 
Other errors can be reduced by increasing the computing
power, and it would be feasible to reduce the total error to
the 5\% level. 

The semi-leptonic decay $D\to\pi\ell\nu$, $K\ell\nu$ form
factors have also been calculated by the same group, and the
results are $f_+^{D\to\pi}=0.64(3)(6)$ and 
$f_+^{D\to K}=0.73(3)(7)$, which can be used to determine
$|V_{cs}|$ and $|V_{cd}|$ with experimental inputs from
BES\cite{Ablikim:2004ej} and CLEO\cite{Liu:2004xr}.
Such determination is currently consistent with the CKM
unitarity and the error is dominated by lattice
calculation. 
If we can reduce the error to the 5\% level or better, it
will provide interesting test of the CKM unitarity.

\subsection{$B$ meson mixings}
The mass difference in the $B^0-\bar{B}^0$ system 
$\Delta M_d$ is proportional to $f_B^2B_B |V_{td}|^2$, and
thus gives a constraint on the CKM element $|V_{td}|$,
provided that the corresponding hadronic matrix elements,
$B$ meson decay constant $f_B$ and $B$ parameter $B_B$, are
theoretically calculated. 
The analogous mass difference in the $B_s^0-\bar{B}_s^0$
mixing $\Delta M_s$ can be used to reduce the theoretical
uncertainty by considering a ratio
$\Delta M_s/\Delta M_d=\xi^2 (M_{B_s}/M_B) |V_{ts}/V_{td}|^2$,
where a ratio 
$\xi\equiv (f_{B_s}B_{B_s}^{1/2})/(f_BB_B^{1/2})$
to describe the SU(3) breaking effect is the quantity to be
estimated theoretically. 

\begin{figure}[tb]
  \centering
  \includegraphics*[width=6.0cm]{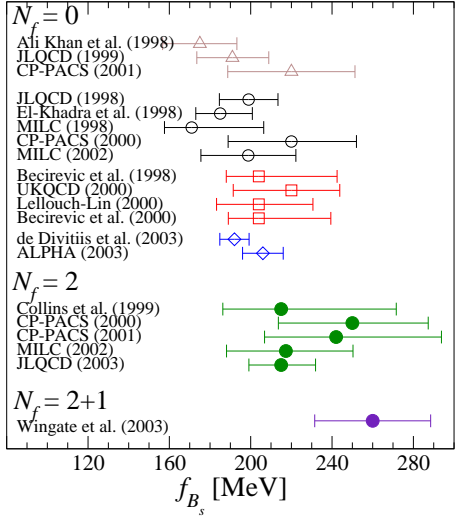}
  \caption{
    Lattice calculations of $f_{B_s}$.
    Quenched ($N_f$ = 0) results are obtained with the NRQCD
    action
    (Ali Khan et al.\protect\cite{AliKhan:1998df},
    JLQCD\protect\cite{Ishikawa:1999xu},
    CP-PACS\protect\cite{AliKhan:2001jg}),
    Fermilab formulation
    (JLQCD\protect\cite{Aoki:1998ji},
    El-Khadra et al.\protect\cite{El-Khadra:1997hq},
    MILC\protect\cite{Bernard:1998xi},
    CP-PACS\protect\cite{AliKhan:2000eg},
    MILC\protect\cite{Bernard:2002pc}),
    and conventional Wilson fermions
    (Becirevic et al.\protect\cite{Becirevic:1998ua},
    UKQCD\protect\cite{Bowler:2000xw},
    Lellouch-Lin\protect\cite{Lellouch:2000tw},
    Becirevic et al.\protect\cite{Becirevic:2000nv},
    de Divitiis et al.\protect\cite{deDivitiis:2003wy},
    ALPHA\protect\cite{Rolf:2003mn}).
    Unquenched ($N_f$ = 2, 2+1) calculations are
    Collins\protect\cite{Collins:1999ff},
    CP-PACS\protect\cite{AliKhan:2000eg,AliKhan:2001jg},
    MILC\protect\cite{Bernard:2002pc},
    JLQCD\protect\cite{Aoki:2003xb},
    Wingate et al.\protect\cite{Wingate:2003gm}.
  }
  \label{fig:fbs}
\end{figure}

In the lattice calculation, the difficulty to treat heavy
quarks on the lattice has been essentially solved by
introducing the HQET based lattice actions, and the results
with different formulations are in good agreement in the
quenched approximation within the systematic uncertainty of
order of 15\%, as shown in Figure~\ref{fig:fbs} for $f_{B_s}$. 

Unquenched simulations have been done since
1999\cite{Collins:1999ff,%
  AliKhan:2000eg,AliKhan:2001jg,Bernard:2002pc}.
Figure~\ref{fig:fbs} shows the results for $f_{B_s}$, which
is relatively insensitive to the problem of chiral log and the
chiral extrapolation is more reliable.
The unquenched results seemed slightly higher than the
quenched results, but it depends on which quantity is used
to set the lattice scale, {\it e.g.} if we set the scale
with the heavy quark potential ($r_0$) the quenched 
and unquenched results are consistent with each other. 
On the other hand, the disagreement between the most recent
two calculations, 
JLQCD ($N_f=2$)\cite{Aoki:2003xb} and 
Wingate {\it et al.} ($N_f=2+1$)\cite{Wingate:2003gm}, 
remains even if one uses the common scale such as $r_0$.
At present, it is not clear if it comes from the dynamical
strange quark effect or from other (underestimated)
systematic errors. 
For an average of unquenched calculations I recommend
$f_{B_s}=230\pm 30$~MeV. 

For the heavy-light meson decay constant, ChPT
predicts a non-analytic quark mass
dependence\cite{Grinstein:1992qt} 
$\propto (1+3g^2) m_\pi^2\ln m_\pi^2$,
whose effect could become important in the chiral
extrapolation of $f_B$.
The coupling $g$ describe the $HH^*\pi$ interaction, such as
$BB^*\pi$ or $DD^*\pi$.
For the $DD^*\pi$ interaction, there is an experimental
measurement $g$ = 0.59(7)\cite{Anastassov:2001cw}, and
quenched lattice
calculations\cite{deDivitiis:1998kj,Abada:2002xe} 
are also consistent with this.

\begin{figure}[tb]
  \centering
  \includegraphics*[width=6.0cm]{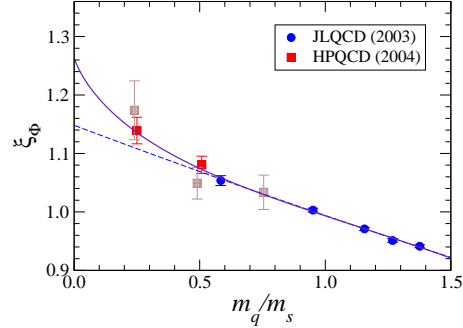}
  \caption{
    A ratio $(f_{B_s}M_{B_s}^{1/2})/(f_BM_B^{1/2})$
    as a function of light and strange quark mass ratio.
    Lattice data are from JLQCD\protect\cite{Aoki:2003xb} 
    and HPQCD\protect\cite{Gray:2004hd}.
  }
  \label{fig:xi}
\end{figure}

In Figure~\ref{fig:xi} the chiral extrapolation is shown for 
a ratio 
$(f_{B_s}M_{B_s}^{1/2})/(f_BM_B^{1/2})$ as a function of
$m_q/m_s$\footnote{
  The plot is an update of that shown by Kronfeld at Lattice 
  2003\cite{Kronfeld:2003sd}.
}.
It goes up toward the physical point $m_q/m_s\approx 1/25$.
The results from the JLQCD collaboration\cite{Aoki:2003xb}
above $m_q/m_s=0.5$ are consistent with a linear dependence,
but the chiral log could raise the chiral limit as shown by
the solid curve. 
Recently, the data from the HPQCD
collaboration\cite{Gray:2004hd} with the staggered sea
quarks become available at smaller quark masses, which are
consistent with the expected chiral log behavior, although
more statistics is needed to be conclusive. 
I fit the both data with a function including the chiral log
and obtain $f_{B_s}/f_B=1.22(^{+5}_{-6})$.

The large uncertainty due to the chiral log can be avoided
by constructing a double ratio\cite{Grinstein:1993ys}
$(f_{B_s}/f_B)/(f_{D_s}/f_D)$, in which the chiral log term
cancels at the leading order of $1/M$.
The JLQCD collaboration calculated this ratio and obtained a
very precise result 1.01(1)\cite{Onogi:2004gd}.
Once the $D$ meson decay constants are measured precisely at
CLEO-c and BES III, this strongly constrains the ratio
$f_{B_s}/f_B$. 
One could also consider another double ratio
$(f_{B_s}/f_B)/(f_K/f_\pi)$, for which the effect
of chiral log is numerically small\cite{Becirevic:2002mh}.

Fortunately, the coefficient of the chiral log term is small
for the $B$ parameter $B_B$, and the unquenched lattice
calculation shows no significant sea quark mass dependence.
Therefore, the results of the JLQCD
collaboration\cite{Aoki:2003xb} will be robust as far as the
chiral extrapolation is concerned. 
Unquenched results are
$B_B(m_b)=0.84(3)(6)$ and $B_{B_s}/B_B=1.02(2)(^{+6}_{-2})$,
which are consistent with the previous quenched
results\cite{Aoki:2002bh}.

\begin{table}[tb]
  \centering
  \begin{tabular}{|c|c|c|}
    \hline
    & Lellouch\cite{Lellouch:2002nj}, & My average \\
    & ICHEP 2002 & ICHEP 2004 \\
    \hline\hline
    $f_B$ & 
    203(27)($^{+\ 0}_{-20}$) & 189(27) \\
    $f_{B_s}$ & 
    238(31) & 230(30) \\
    $f_B\hat{B}_B^{1/2}$ & 
    235(33)($^{+\ 0}_{-24}$) & 214(38) \\
    $f_{B_s}\hat{B}_{B_s}^{1/2}$ & 
    276(38) & 262(35) \\
    $f_{B_s}/f_B$ &
    1.18(4)($^{+12}_{-\ 0}$) & 1.22($^{+5}_{-6}$) \\
    $\xi$ &
    1.18(4)($^{+12}_{-\ 0}$) & 1.23(6) \\
    \hline
  \end{tabular}
  \caption{
    My averages of $B$ mixing parameters compared to those
    by Lellouch at ICHEP 2002\protect\cite{Lellouch:2002nj}.
    $f_{B_{(s)}}$ and 
    $f_{B_{(s)}}\hat{B}_{B_{(s)}}^{1/2}$
    are given in units of MeV.
  }
  \label{tab:summary}
\end{table}

My averages for the $B-\bar{B}$ mixing parameters are
summarized in Table~\ref{tab:summary}.
The averages at ICHEP 2002 provided by
Lellouch\cite{Lellouch:2002nj} are also listed for
comparison.
The main difference is the treatment of the uncertainty due
to the chiral extrapolation.
Because an indication of the chiral log is already found I
take the central value from the fit including the chiral
log, while it was given in the asymmetric
second error in \cite{Lellouch:2002nj}.

\subsection{$B$ meson decays}
The $B$ meson semi-leptonic decay form factors are needed in
the determination of $|V_{ub}|$ and $|V_{cb}|$ from
exclusive decay modes $B\to\pi\ell\nu$ and $B\to D^{(*)}\ell\nu$.

\begin{figure}[tb]
  \centering
  \includegraphics*[width=6.5cm]{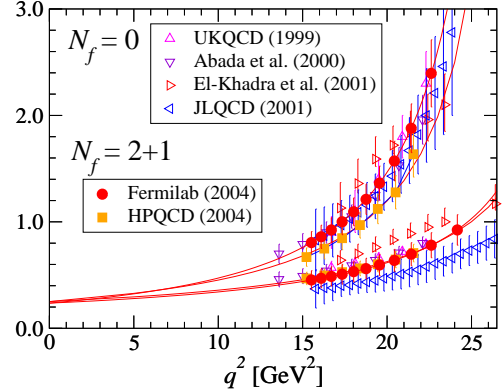}
  \caption{
    The $B\to\pi\ell\nu$ decay form factors $f_+(q^2)$
    (upper cluster of points) and $f_0(q^2)$ (lower cluster)
    from quenched and unquenched lattice QCD.
    Quenched ($N_f=0$) results are from
    UKQCD\protect\cite{Bowler:1999xn},
    Abada et al.\protect\cite{Abada:2000ty},
    El-Khadra et al.\protect\cite{El-Khadra:2001rv},
    JLQCD\protect\cite{Aoki:2001rd}, while the unquenched results
    are from
    Fermilab\protect\cite{Okamoto:2004xg} and 
    HPQCD\protect\cite{Shigemitsu:2004ft}.
  }
  \label{fig:B2pi}
\end{figure}

The first unquenched calculations of the
$B\to\pi\ell\nu$ form factors have recently been presented
by the Fermilab lattice\cite{Okamoto:2004xg}
and the HPQCD\cite{Shigemitsu:2004ft} collaborations, 
which both worked on 2+1-flavor gauge field ensembles produced
by the MILC collaboration.
Figure~\ref{fig:B2pi} shows their results together with the
previous quenched
calculations\cite{Bowler:1999xn,Abada:2000ty,El-Khadra:2001rv,Aoki:2001rd}.
We do not observe any significant difference
between quenched and unquenched calculations within
relatively large statistical errors.

Since lattice calculations are possible only in the high
$q^2$ region where the recoil
momentum of the daughter pion is small, the extraction of
$|V_{ub}|$ using these lattice results requires 
experimental data in the same kinematical region.
Such analysis was done by the CLEO
collaboration\cite{Athar:2003yg}.
They provided the data of the decay rate in three $q^2$
bins, and the highest bin (above 16~GeV$^2$) corresponds
to the region of the lattice calculation.
The result using an average of the four quenched lattice
calculations
$|V_{ub}|=(2.88\pm 0.55\pm 0.30\pm^{+0.45}_{-0.35}\pm
0.18)\times 10^{-3}$
still has large errors (due to statistical, systematic,
lattice, and $\rho\ell\nu$ form factor dependence, in the
order given), but we expect better measurements from BaBar
and Belle in near future. 
In fact, a preliminary result with higher purity (and thus
with smaller experimental systematic error) was reported at
this conference by the Belle
collaboration\cite{Abe:2004zm}.

The Fermilab lattice collaboration has
updated\cite{Okamoto:2004xg} the 
calculation of $B\to D\ell\nu$ form factor at zero recoil,
which can be used in the precise determination of
$|V_{cb}|$. 
Their new result with 2+1 flavors of dynamical quarks,
${\cal F}_{B\to D}(1)=1.075(18)(15)$,
is consistent with the previous quenched
calculation\cite{Hashimoto:1999yp}.
Similar calculation of the $B\to D^{(*)}\ell\nu$ form factor
\cite{Hashimoto:2001nb} is in progress.

\section{Conclusions}
The most important progress in lattice QCD calculations in
the past few years is the inclusion of sea quark effects.
By performing the real simulations with dynamical quarks,
the {\it unknown} errors due to quenching 
in the previous calculations are being eliminated for many
of the quantities discussed here.

In unquenched calculations, the non-analytic quark mass 
dependence appearing from the pion loops must be
taken into account in the chiral extrapolation.
Without enough data points below $m_{PS}\simeq$ 500~MeV, the
extrapolation induces large systematic uncertainty, since we
do not know where the chiral log effect becomes important.
In this respect the advantage of staggered
fermion is clear: the simulation is the fastest and
therefore one can reach the chiral regime.
Employing the improved staggered fermions, the
HPQCD-MILC-UKQCD-Fermilab group have recently presented
several results with 2+1 flavors of dynamical quarks.

The drawback of the staggered fermion is its complicated
taste structure, in particular the fourth-root trick
introduced to represent single dynamical flavor.
It is not proved that the fourth-root of the staggered
fermion determinant is the same as a local quantum field
theory.
Such a proof is essential, as it would provide a theoretical
basis that these calculations correspond to the real QCD.
Until such a proof is available, the simulation with other
fermion formulations without such taste structure should be
pursued.

\begin{figure}[tb]
  \centering
  \includegraphics*[width=7.5cm]{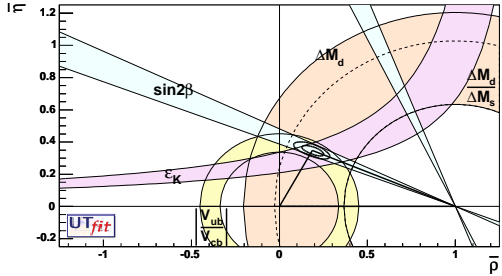}
  \caption{
    Constraints on the CKM unitarity triangle using the
    most recent lattice results for $B_K$,
    $f_{B_s}B_{B_s}^{1/2}$, and $\xi$.
    Plot is from the UTfit collaboration\protect\cite{Bona:2004sj}.
  }
  \label{fig:CKM}
\end{figure}

The constraints on the CKM unitarity triangle using the
recent lattice results are
shown in Figure~\ref{fig:CKM}.
Although the error bars are not significantly reduced over
the last several years, the uncertainty due to quenching is
now basically eliminated, and the calculations are from 
{\it the} first principles.
Much more work is needed to reduce other systematic errors,
and the ideas are already being tested within the quenched
approximation.

\section*{Acknowledgments}
I would like to thank 
I. Allison, Y. Aoki, C. Bernard, N. Christ, C. Dawson,
J. Flynn, E. Gamiz, A. Gray, R. Horsley, T. Iijima,
T. Izubuchi, J. Laiho, C.J.D. Lin, Q. Mason,
C. Maynard, F. Mescia, J. Noaki, M. Okamoto, C. Pena,
M. Pierini, G. Schierholz, J. Shigemitsu, J. Simone,
A. Soni, A. Stocchi, S. Tamhanker, M. Wingate, H. Wittig, 
and the members of the CP-PACS and JLQCD collaborations 
for correspondence and discussions.
I apologize that some of their works could not be included
in this review. 
I also thank Andreas Kronfeld for carefully reading the
manuscript.
This work is supported in part by the Grant-in-Aid of the
Ministry of Education (No.~14540289).

\end{document}